\documentclass[prx,twocolumn,showpacs,amsmath,amssymb,superscriptaddress]{revtex4-2}
\usepackage{braket}
\usepackage{cases}
\usepackage{color}
\usepackage{graphicx,subfigure}
\usepackage{float}
\usepackage{bm}
\usepackage{hyperref}
\hypersetup{
    colorlinks=true, 
    linkcolor=cyan,
    citecolor=magenta, 
    filecolor=magenta, 
    urlcolor=cyan,
    runcolor=cyan
}
\usepackage[capitalise]{cleveref}

\newcommand{\eps} {\varepsilon}
\newcommand{\epstot} {\eps_{\mathrm{tot}}}
\newcommand{\epserr} {\eps_{\mathrm{err}}}
\newcommand{\delerr} {\Delta_{\mathrm{err}}}

\newcommand{\dphi}{\delta\varphi}
\newcommand{\dth}{\delta\theta}
\newcommand{\Herr}{H_{\mathrm{err}}}

\usepackage{commath}
\usepackage{color}
\usepackage{amsmath}
\usepackage{adjustbox}
\usepackage[normalem]{ulem}

\newcommand{\bes} {\begin{subequations}}
\newcommand{\ees} {\end{subequations}}
\newcommand{\beq}{\begin{equation}}
\newcommand{\eeq}{\end{equation}}

\def\>{\rangle}
\def\<{\langle}

\newcommand{\VZ}{\textsc{VZ} }
\def\XY4{\text{XY}4}

\begin{document}

\title{Virtual $Z$ gates and symmetric gate compilation}

\author{Arian Vezvaee}
\thanks{These two authors contributed equally to this work.}
\affiliation{Department of Electrical \& Computer  Engineering, University of Southern California, Los Angeles, California 90089, USA}
\affiliation{Center for Quantum Information Science \& Technology, University of
Southern California, Los Angeles, CA 90089, USA}

\author{Vinay Tripathi}
\thanks{These two authors contributed equally to this work.}
\affiliation{Center for Quantum Information Science \& Technology, University of
Southern California, Los Angeles, CA 90089, USA}
\affiliation{Department of Physics \& Astronomy, University of Southern California,
Los Angeles, California 90089, USA}

\author{Daria Kowsari}
\affiliation{Center for Quantum Information Science \& Technology, University of
Southern California, Los Angeles, CA 90089, USA}
\affiliation{Department of Physics \& Astronomy, University of Southern California,
Los Angeles, California 90089, USA}

\author{Eli Levenson-Falk}
\affiliation{Department of Electrical \& Computer  Engineering,  University of Southern California, Los Angeles, California 90089, USA}
\affiliation{Center for Quantum Information Science \& Technology, University of
Southern California, Los Angeles, CA 90089, USA}
\affiliation{Department of Physics \& Astronomy, University of Southern California,
Los Angeles, California 90089, USA}

\author{Daniel A. Lidar}
\affiliation{Department of Electrical \& Computer  Engineering,  University of Southern California, Los Angeles, California 90089, USA}
\affiliation{Center for Quantum Information Science \& Technology, University of
Southern California, Los Angeles, CA 90089, USA}
\affiliation{Department of Physics \& Astronomy, University of Southern California,
Los Angeles, California 90089, USA}
\affiliation{Department of Chemistry, University of Southern California,
Los Angeles, California 90089, USA}

\begin{abstract}
The virtual $Z$ gate has been established as an important tool for performing quantum gates on various platforms, including but not limited to superconducting systems. Many such platforms offer a limited set of calibrated gates and compile all other gates using combinations of $X$-type and virtual $Z$ gates.  Here, we show that the method of compilation has important consequences in an open quantum system setting. Specifically, we experimentally demonstrate that it is crucial to choose a compilation that is symmetric with respect to virtual $Z$ rotations. An important example is dynamical decoupling (DD) sequences, where improper gate decomposition can result in unintended effects such as the implementation of the wrong sequence. Our findings indicate that in many cases the performance of DD is adversely affected by the incorrect use of virtual $Z$ gates, compounding other coherent pulse errors. This holds even for DD sequences designed to be robust against systematic control errors. 
In addition, we identify another source of coherent errors: interference between consecutive pulses that follow each other too closely. This work provides insights into improving general quantum gate performance and optimizing DD sequences in particular.

\end{abstract}

\maketitle


\section{Introduction}
Any quantum computing processor is inherently an open quantum system that interacts with its environment, leading to decoherence and errors, which adversely affect quantum computations~\cite{nielsen2010quantum}. Various error correction, suppression, and mitigation techniques are employed to suppress these effects~\cite{LidarBrun2013QEC,alvarez2016MRP,Campbell:2017aa,Cai2023RMP}. There has been a great interest in demonstrations of overcoming decoherence, which have recently become possible with the availability of commercial cloud-based quantum processors~\cite{mell2011nist,IBMQ,Karalekas2020QST,wurtz2023arxiv,Blinov2021AZS}. These quantum processors usually have a native set of calibrated gates from which all other gates can be constructed. An important part of the native gate set is the Virtual-$Z$ (\textsc{VZ}) gate, which is an instantaneous, error-free operation that plays a central role in gate compilation. Ref.~\cite{McKay2017PRA} demonstrated that \VZ gates can be implemented by simply adding a phase offset in software, unlike physical $Z$-gates that involve physical rotations around the $z$-axis of the Bloch sphere. Moreover, they showed, by manipulating the phases of pulses driving the qubits, \VZ gates can be effectively combined with two $\sqrt{X}$ gates to construct any SU(2) gate, thus achieving universality when combined with a two-qubit entangling gate~\cite{Lloyd:95,Barenco:1995aa}. This approach simplifies the gate decomposition and circuit compilation procedure, and its applicability extends beyond qubits to qudits as well as beyond superconducting systems~\cite{Liu2023PRX,Vezvaee2023PRX,GullansPRXQ2024,Tsai2022PRApp,Fischer2022PRR,vezvaee2024arxiv,KazminaPRA2024}. Compiling an arbitrary SU(2) operation using \VZ gates provides flexibility, but ensuring accuracy in the presence of open quantum system effects is essential for reliable computations. Although different compilations involving \VZ gates can be equivalent in closed systems, discrepancies may arise if open-system effects are not correctly accounted for during compilation.

In this work, we investigate the role of \VZ gates in gate compilation within an open quantum system dynamics framework.  We find that even slight variations in the compilation of quantum gates using \VZ gates has significant impact. Specifically, an asymmetric compilation relative to \VZ gates of any gate corresponding to a rotation about an axis in the $(x,y)$ plane leads to noticeable effects in various settings, including dynamical decoupling (DD) and algorithm execution. For example, asymmetric compilation of the $Y$ gate introduces fidelity discrepancies between the $Y$ eigenstates $\ket{\pm i}$, which can be completely mitigated with proper compilation techniques. This observation has important consequences, which we explore in this work. 

Asymmetric compilation {affects} the implementation of DD sequences~\cite{Viola:98,Viola1998PRL,Zanardi:1999fk,Duan:98e,Vitali:99,Viola:02,Viola:2005:060502,Khodjasteh:2005xu,Uhrig:2007qf,Souza:2011aa,Quiroz:2013fv,Genov2017PRL,Pokharel2018PRL,EzzellPRApp2021,GustavssonPRL2013}, potentially leading to misimplementation and misidentification of commonly used sequences. For example, DD implementations using cloud quantum processors reveal unexpected pulse-interval effects, as well as unexplained significant oscillations in single-qubit experiments~\cite{EzzellPRApp2021}. In addition, our findings uncover previously unrecognized oscillations, even in DD sequences designed to be robust to coherent errors~\cite{Souza:2011aa,Quiroz:2013fv,Genov2017PRL}. 
The investigations into the \VZ gate we report here reveal that interference between consecutive pulses explains these oscillations in robust sequences. Given the recent critical role DD has played in improving the fidelity of quantum states~\cite{Pokharel2018PRL,Souza_2021,Tripathi2022,EzzellPRApp2021,tong2024arxiv,seif2024arxiv,Rahman2024PRApp}, circuits~\cite{Arute2019Nature,jurcevicDemonstrationQuantumVolume2021,baumer2023efficientlongrangeentanglementusing}, and even entire algorithms~\cite{Pokharel2023PRL,Pokharel2024npj,Singkanipa2024arxiv,baumer2024quantumfouriertransformusing}, we expect these findings to contribute to further improvement of quantum error suppression via pulse-based methods such as DD. However, the impact extends beyond DD to any quantum algorithm or error-correction method that requires high-fidelity single-qubit gates. 


The structure of this paper is as follows. In \cref{sec:VZ-open-sys} we provide the theoretical background for the difference between symmetric and asymmetric gate compilation in an open system setting. This includes a discussion of coherent error sources, an intuitive trajectory-based picture for symmetric \textit{vs} asymmetric compilation, and the impact on DD. \cref{sec:expt} contains all our experimental results, obtained using our in-house $\texttt{MUNINN}$ processor as well as IBMQ devices. We present results for various DD sequences, GHZ state fidelity, 
and analyze pulse interference effects. Our result present strong confirmation of the advantage offered by symmetric gate compilation. We conclude in \cref{sec:conc}.

\begin{figure*}
\hspace{0cm}{\includegraphics[scale=.6]{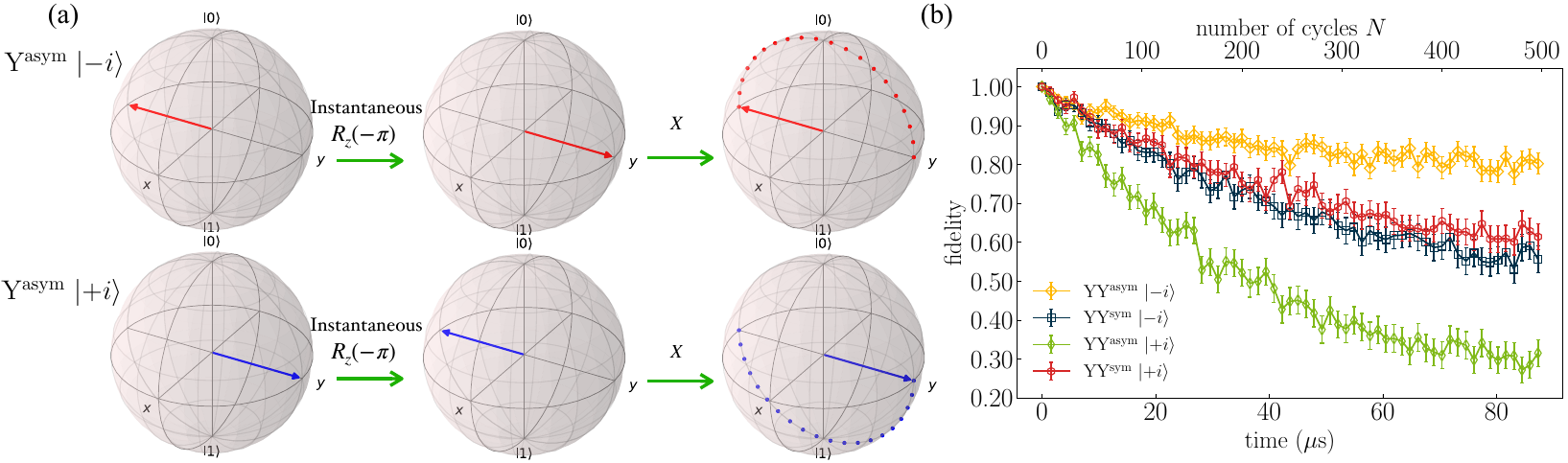}} 
\caption{The open system effect of the symmetric and asymmetric compilation of the $Y$ gate with respect to \VZ gates. (a) The $\ket{\pm i}$ state follows different Bloch sphere trajectories under $Y^\text{asym}$, which consists of an instantaneous \VZ gate followed by a physical $X$ gate. This causes $\ket{-i}$ ($\ket{+i}$) to go through a stable (unstable) ground (excited) state which leads to the asymmetry in the fidelity of the two states. The symmetric decomposition $Y^\text{sym}$ overcomes this asymmetry, similar to a physical $Y$ gate. (b) Experimental demonstration of the symmetric and asymmetric effects of the $Y$-gate decomposition on the \texttt{MUNINN} processor. The fidelity of the states $\ket{\pm i}$ is shown under both $Y^\text{asym}$ and $Y^\text{sym}$, as a function of time (bottom axis) or number of $YY$ sequence cycles (top axis). The symmetric decomposition results in similar fidelities (black and red) for the initial states $\ket{\pm i}$. The asymmetric decomposition results in very different fidelities (yellow and green) for the same two initial states{, in agreement with our theoretical prediction of a higher fidelity for the $\ket{-i}$ state, whose trajectory passes through the ground state $\ket{0}$}. Here and in all other figures, error bars denote two standard deviations of the mean.} 
\label{fig:one}
\end{figure*}

\section{\VZ gate in an open quantum system}
\label{sec:VZ-open-sys}

We conduct all our experiments using two superconducting transmon quantum processors: the IBM cloud quantum processor \texttt{ibm\_sherbrooke} and our in-house quantum processor $\texttt{MUNINN}$~\cite{muninn2024}; the design can be found in the SQuADDS database \cite{shantoSQuADDSValidatedDesign2024}.

We model the transmon qubit as a driven two-level system and consider it in the drive frame under the rotating wave approximation (e.g., Ref.~\cite{TripathiTransmon2024}). Let $\{\sigma_\alpha\}$ denote the set of Pauli matrices. The time-dependent system Hamiltonian that generates single-qubit $X$ rotation gates is given by:
\bes
    \label{eq:Hsys}
\begin{align}
        H(t) &= \epstot(t) \frac{\sigma_x}{2} + \Herr  \\
        \Herr &=  \epserr \frac{\sigma_x}{2} + \delerr  \frac{\sigma_z}{2} .
\end{align}
\ees
Here, $\epstot(t)$ is the intended time-dependent control field and $\epserr$ and $\delerr$ are errors. Ideally, $\epserr = \delerr  = 0$. In reality, both are present and give rise to rotation and phase errors
{
\begin{equation}
\dth\equiv \epserr  t_g\ , \quad  \dphi \equiv \frac{\delerr}{\bar{\eps}} = \frac{1}{\theta}\delerr t_g
\label{eq:dth-dphi}
\end{equation}
respectively, with $t_g$ denoting the gate duration, $\theta \equiv  \int_0^{t_g}\epstot(t)dt$, and $\bar{\varepsilon} = \theta/t_g$ the 
average pulse amplitude~\cite{tripathiBenchmarkingQuantumGates2025}.}
An open system single-qubit gate includes both rotation and phase errors, as well as a system-bath interaction term that is always present while the gate is being generated.

\subsection{Gate compilation}

\VZ gates eliminate the need for performing physical rotations about the Bloch $z$-axis, allowing us to focus solely on rotations in the $(x,y)$ plane.  We denote by $R_{\phi}(\theta)$ a rotation by an angle $\theta$ about an arbitrary axis in the $(x,y)$ plane, making an angle $\phi$ with the $x$-axis: 
\beq
\label{eq:r-phi}
R_{\phi}(\theta) \equiv \exp[-i (\theta/2) (\cos(\phi)\sigma_x + \sin(\phi)\sigma_y)] .
\eeq
We also denote 
\beq
R_x(\theta) \equiv R_{\phi=0}(\theta) \ , \quad R_y(\theta) \equiv R_{\phi=\pi/2}(\theta) .
\eeq
The physical implementation of $R_{\phi}(\theta)$ involves applying an on-resonance microwave pulse of the form $\epstot(t) = \varepsilon(t) \cos(\omega t+\phi)$ to the qubit, where the integrated pulse amplitude (for a given pulse duration) determines $\theta$, and the pulse phase determines $\phi$. The phase $\phi$ is arbitrary, as is the choice of the $(x,y)$ coordinate system, both set by the initial pulse. This illustrates how the \VZ gate is implemented simply by updating the definition of which pulse phase corresponds to $\phi = 0$ (usually set to be the $x$-axis, as above). However, this adjustment has tangible physical effects on subsequent gates: after a virtual $R_z(\varphi) \equiv \exp[-i(\varphi/2)\sigma_z]$ gate [note that $R_z(\pi) \equiv Z =-i\sigma_z$], the phase of each of the rotations that follow is shifted by $\varphi$. For example, when $\varphi=\pi$, then the next operation $R_x(\theta)$ becomes $R_{\pi}(\theta) = R_{-x}(\theta)$, i.e., a rotation about the $x$-axis becomes a rotation about the $-x$ axis, in the sense that $R_{-x}(\theta) = R^\dag_z(\pi)R_x(\theta)R_z(\pi)$.

For most commercial cloud-based quantum processors not all rotations are natively available. For example, for the IBMQ devices, the calibrated single-qubit native gate set typically consists of the operations $\mathcal{G}=\{R_z(\varphi), \sqrt{X}, X\}$, where $\sqrt{X}\equiv R_x(\pi/2)$ and $X \equiv R_x(\pi) = -i \sigma_x$, which are generated using $H(t)$ given in \cref{eq:Hsys}. However, these are not the only Clifford operations necessary for universal quantum computation. All other Clifford gates must be decomposed into these operations. Specifically, a $Y\equiv R_y(\pi) = -i\sigma_y$ gate requires an $X$ gate combined with \VZ gates, which can be done in different ways. One method of compiling a $Y$ gate is {asymmetric}:
\begin{equation}
\label{eq:Yasym}
Y^{\text{asym}}=XR_z(-\pi).
\end{equation} 
Alternatively, a symmetric compilation of the $Y$ gate with respect to the \VZ gates is:
\begin{equation}
\label{eq:Ysym}
Y^{\text{sym}}=R_z(\pi/2)XR_z(-\pi/2).
\end{equation}
Although these methods are theoretically equivalent in the sense that $Y^{\text{asym}}=Y^{\text{sym}}=Y$ is a mathematical identity, this is no longer the case when one accounts for deviations from unitary dynamics due to open quantum system effects, as we discuss in detail below.

\subsection{Trajectories matter: asymmetry between $\ket{+i}$ and $\ket{-i}$}

To demonstrate how the two compilation strategies result in different outcomes, we consider a simple experiment, in which we apply sequences with a varying number of $YY$ pulses to the two orthogonal initial states $\ket{\pm i}$. Ideally, the fidelity of $YY$ applied to $\ket{+i}$ or $\ket{-i}$ should be identical. However, with the asymmetric decomposition the two states follow different Bloch sphere trajectories and leave the $(x,y)$ plane. That is, in the case of $Y^\text{asym}$, the virtual $R_z(-\pi)$ gate instantaneously interchanges $\ket{+i}$ and $\ket{-i}$ (up to a global phase of $i$) before the physical $X$ gate is applied. This has the effect of $\ket{-i}$ following a trajectory through the stable ground state $\ket{0}$ during the $X$ gate, while $\ket{+i}$ passes through the unstable excited state $\ket{1}$ [see \cref{fig:one}(a)]. The second $Y$ gate leads to a reversal of this trajectory, again passing through the ground/excited, as the virtual $R_z(-\pi)$ reverses the direction of rotation. Consequently, $\ket{-i}$ experiences a lower relaxation rate and maintains a higher fidelity compared to $\ket{+i}$ over the course of repeated applications of the $YY$ sequence. 

Conversely, using the symmetric decomposition $Y^\text{sym}$, the first \VZ gate, $R_z(-\pi/2)$ transforms  $\ket{-i}$ to $\ket{-}$ and $\ket{+i}$ to $\ket{+}$ (up to a global phase of $e^{i \pi/4}$). These states then undergo an $X$ gate, which leaves them unchanged (up to a global phase). The next \VZ gate, $R_z(\pi/2)$, transforms the state back to its original position on the $y$-axis. Therefore, with this compilation, $\ket{\pm i}$ both remain in the $(x,y)$ plane at all times during the $Y^\text{sym}$ gate, and do not experience different relaxation rates. As a result, the fidelities of $\ket{\pm i}$ should be similar under $YY$, as for a physical $Y$ gate. By linearity, this extends to any state in the $(x,y)$ plane, i.e., to any superposition of $\ket{\pm i}$ or $\ket{\pm}$. Another way to see this result is that symmetric $YY$ compiles to two repetitions of \cref{eq:Ysym}, which is then equal to $R_z(\pi/2)XXR_z(-\pi/2)$. The interior $XX$ sequence traces out a full $2\pi$ rotation and thus always leads to trajectories that trace the same paths for any two opposite initial states, as expected by the $YY$ sequence.

\begin{figure}[ht]
\hspace{0cm}{\includegraphics[scale=.43]{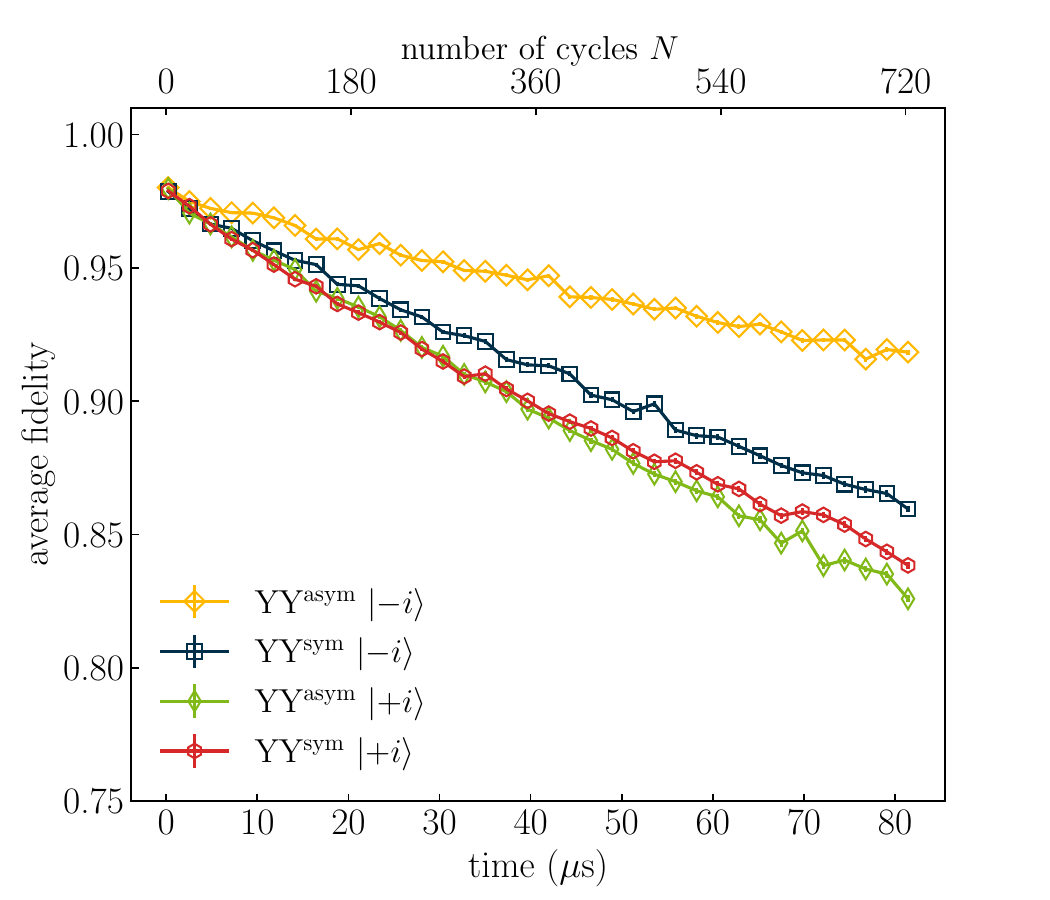}}
\caption{Effect of the asymmetric $Y^{\text{asym}}$ and symmetric $Y^\text{sym}$ gates on state fidelity averaged over four qubits ($0,13,81,89$) of the \texttt{ibm\_sherbrooke} processor. The fidelity is shown as a function of time (bottom) or the number of cycles of the $YY$ sequence (top). We observe that subject to $Y^\text{sym}$ the fidelities of the $\ket{\pm i}$ states (black and red) are significantly closer than subject to $Y^\text{asym}$ (yellow and green).} 
\label{fig:two}
\end{figure}

We verified the effects predicted above through various experiments, using both the \texttt{ibm\_sherbrooke} and  $\texttt{MUNINN}$ processor. As described above, we first prepare the initial states $\ket{\pm i}$, apply a series of $YY$ sequences, unprepare the initial state, and measure the system in the $\sigma_z$ eigenbasis. We define the empirical fidelity as the frequency of favorable outcomes, i.e., the number of $\ket{0}$ outcomes divided by the total number of experimental shots ($800$). \cref{fig:one}(b) shows the results on the $\texttt{MUNINN}$ processor, where we demonstrate that the asymmetric compilation of the $Y$ gate leads to the predicted asymmetry in the fidelity of the $\ket{\pm i}$ states. In contrast, the two states decay almost identically when the symmetric compilation of $Y$ gate is used. We observe the same effect after repeating the experiments on different qubits of the \texttt{ibm\_sherbrooke} processor, as highlighted in \cref{fig:two}.

\subsection{Impact on DD sequences}

Next, we consider the impact of symmetric \textit{vs} asymmetric compilation on DD sequences.

\subsubsection{Correct $\overline{X} \equiv R_x(-\pi)$}

As discussed earlier, the \VZ gates enact a frame transformation for all subsequent gates. In the present context, this manifests as the identity
\beq
\label{eq:5}
R_x(\pi) R_z(\pm \pi) = - R_z(\mp \pi) R_x(-\pi) ,
\eeq
The frame transformation defined by \cref{eq:5} can be reinterpreted as a way to create a correct $\overline{X}\equiv R_x(-\pi)$ gate, which plays an important role in robust DD sequences~\cite{Ng:2011dn,Quiroz:2013fv,Genov2017PRL}. Namely, we perform the symmetric version of the gate as:
\begin{equation} 
\label{eq:xbar}
    \overline{X} = R_z(-\pi) X R_z(\pi) \text{ or } R_z(\pi) X R_z(-\pi) .
\end{equation}
As we show in {\cref{sec:expt}}, performing the correct $\overline{X}$ gate is critical for understanding the oscillations in the fidelity of the robust DD sequences that have been observed on IBM devices~\cite{EzzellPRApp2021}. In particular, it is essential that the implemented physical rotation is $R_x(-\pi)$ as opposed to $R_x(\pi)$ along with subsequent frame updates.

\subsubsection{Symmetric $Y$ yields XY4}

One of the fundamental DD sequences is XY4~\cite{Maudsley1986ty}:
\begin{equation}
\label{eq:XY4}
    \XY4\equiv Y f_\tau X f_\tau Y f_\tau X f_\tau ,
\end{equation}
where $f_\tau = e^{-i\tau H}$ denotes the free evolution unitary generated by the total system-bath Hamiltonian $H$, and $\tau$ is the pulse interval, i.e., the delay  between two consecutive pulses. XY4 is also known as the universal decoupling sequence, as it decouples arbitrary single-qubit decoherence~\cite{Viola1998PRL}.

We now demonstrate that symmetric compilation yields XY4. Toward this end, first note the identity
\beq
R_x(\pi)R_z(-\pi/2) = R_z(-\pi/2)R_y(\pi) ,
\eeq
i.e., $R_z(-\pi/2)$ changes the rotation axis of the subsequent gates by $\pi/2$, transforming $R_x(\pi)\to R_y(\pi)$. Therefore, using $Y^{\text{sym}}$ [\cref{eq:Ysym}], we have:
\begin{align} 
  \XY4^{\text{sym}} = 
  &\ R_z(\pi/2)  R_x(\pi) R_z(-\pi/2)f_\tau R_x(\pi) f_\tau \times \notag \\
  &\ R_z(\pi/2)  R_x(\pi) R_z(-\pi/2)f_\tau R_x(\pi) f_{\tau} \notag\\
 =&\ R_z(\pi/2) R_z(-\pi/2) R_y(\pi) f_\tau R_x(\pi) f_\tau \times \notag \\
  &\ R_z(\pi/2)  R_z(-\pi/2) R_y(\pi) f_\tau R_x(\pi) f_{\tau} \notag\\
= &\ R_y(\pi)f_\tau R_x(\pi) f_\tau R_y(\pi)f_\tau R_x(\pi) f_{\tau} ,
    \label{eq:xy4-actual}
\end{align}
which is indeed the XY4 sequence given in \cref{eq:XY4}.

\subsubsection{Asymmetric $Y$ yields UR$_4$ instead of XY4}

Using the asymmetric definition of the $Y$ gate [\cref{eq:Yasym}], the XY4 sequence becomes:
\begin{align}
    &\XY4^{\text{asym}} =  \\
    &\quad R_x(\pi) R_z(-\pi)f_\tau R_x(\pi) f_\tau R_x(\pi) R_z(-\pi)f_\tau R_x(\pi) f_{\tau}. \notag
\end{align}
Using \cref{eq:5} allows us to commute the \VZ gate to the left. Since it is a virtual gate implemented via phase offsets in software, the \VZ gate commutes with the free evolution operator $f_\tau$. 
Dropping overall phase factors, we can thus rewrite $\XY4^{\text{asym}}$ as follows:
\begin{align}
    &\XY4^{\text{asym}} =  \notag \\
&\quad R_x(\pi) R_z(-\pi)f_\tau R_x(\pi) R_z(\pi) f_\tau R_x(-\pi)f_\tau R_x(\pi) f_{\tau} = \notag\\
&\quad R_x(\pi) R_z(-2\pi)f_\tau R_x(-\pi) f_\tau R_x(-\pi)f_\tau R_x(\pi) f_{\tau} = \notag\\
& \quad R_x(\pi) f_\tau R_x(-\pi) f_\tau R_x(-\pi) f_\tau R_x(\pi) f_{\tau} . 
\end{align}  
In fact, this sequence is the fourth order ``universally robust''  sequence~\cite{Genov2017PRL}, 
\beq
\label{eq:UR4}
\text{UR}_4 = X f_\tau\overline{X} f_\tau \overline{X} f_\tau X f_\tau ,
\eeq
rather than the intended XY4. I.e.,
\beq
\XY4^{\text{asym}} = \text{UR}_4 .
\eeq

\subsubsection{Generalization to UR$_n$}

UR$_4$ is the lowest order sequence in the UR$_n$ family, defined as:
\bes
\begin{align}
\mathrm{UR}_n &= R_{\phi_n}(\pi)f_\tau\cdots R_{\phi_2}(\pi)f_\tau R_{\phi_1}(\pi)f_\tau, \\
\phi_k &=\frac{(k-1)(k-2)}{2} \Phi^{(n)}+(k-1) \phi_2, \\
\Phi^{(4 m)} &=\frac{\pi}{m} \Phi^{(4 m+2)}=\frac{2 m \pi}{2 m+1},
\end{align}
\ees
where $R_{\phi}(\pi)$ is a $\pi$ rotation about an axis which makes an angle $\phi$ with the $x$-axis, as defined in \cref{eq:r-phi}. 
The phase $\phi_1$ is set to zero without loss of generality. The phase $\phi_2$ can be chosen at will and when it is set to $\pi/2$ for $n=4$ once recovers the XY4 sequence, i.e., $\XY4^{\text{sym}}$. Following \cite{Genov2017PRL}, we set $\phi_2=\pi$, yielding UR$_4$.

Theory predicts that the UR$_n$ family of sequences is robust against systematic rotation and phase errors [\cref{eq:dth-dphi}], i.e., an infidelity that decreases exponentially in the sequence order $n$ at a cost that grows only linearly in $n$~\cite{Genov2017PRL}. In this sense, the fact that asymmetric compilation yields UR$_4$ rather than XY4 can be viewed as an advantage if the goal is only to suppress dephasing. However, unlike the universal XY4 sequence, UR$_4$ cannot suppress multi-axis errors. From this perspective, asymmetric compilation has resulted in a loss of multi-axis suppression capabilities. Note also that since XY4 is a member of the UR$_n$ family (as noted above), it is expected to exhibit the same robustness properties as UR$_4$.

Nevertheless, since it amounts to a type of inherent fault tolerance, there is significant interest in demonstrating the improvement in systematic error suppression associated with the UR$_n$ family. This requires a correct implementation of $R_{\phi}(\pi)$, i.e., a rotation about an arbitrary axis in the $(x,y)$ plane.

While in principle one can write the required arbitrary axis-rotations as
\begin{equation}
\label{eq:rphi-asym}
    R_{\phi}^{\rm{asym}}(\pi)\equiv X R_z(-2\phi),
\end{equation}
only the symmetric choice yields the correct universally robust sequence:
\begin{equation}
\label{eq:rphi-sym}
    R_{\phi}^{\rm{sym}}(\pi)\equiv R_z(-\phi) X R_z(\phi).
\end{equation}
This is the most general form of a $\pi$ rotation about an arbitrary axis in  the $(x,y)$ plane. In fact, \cref{eq:Ysym,eq:xbar} are the special cases of $\phi=\pi/2$ and $\phi=\pi$, respectively. We demonstrate the performance impact on UR$_n$ of the symmetric and asymmetric versions of $R_{\phi}(\pi)$ in \cref{sec:expt-DD,sec:GHZ-expt}.

\subsection{Quantum circuits using symmetric \textit{vs} asymmetric compilation}

As we just saw, the choice between symmetric or asymmetric compilation is not limited to $Y$ or $\overline{X}$ gates, but applies to any axis of rotation in the $(x,y)$ plane. This has significant implications beyond DD and is relevant to a broad range of quantum algorithms. 
We demonstrate the effect of symmetric \textit{vs} asymmetric compilation on different types of quantum circuits in \cref{sec:algos}. These demonstrations contrast circuits employing symmetric compilation [\cref{eq:rphi-sym}] instead of its asymmetric counterpart [\cref{eq:rphi-asym}].

\begin{figure*}[t]
\hspace{0cm}{\includegraphics[scale=.65]{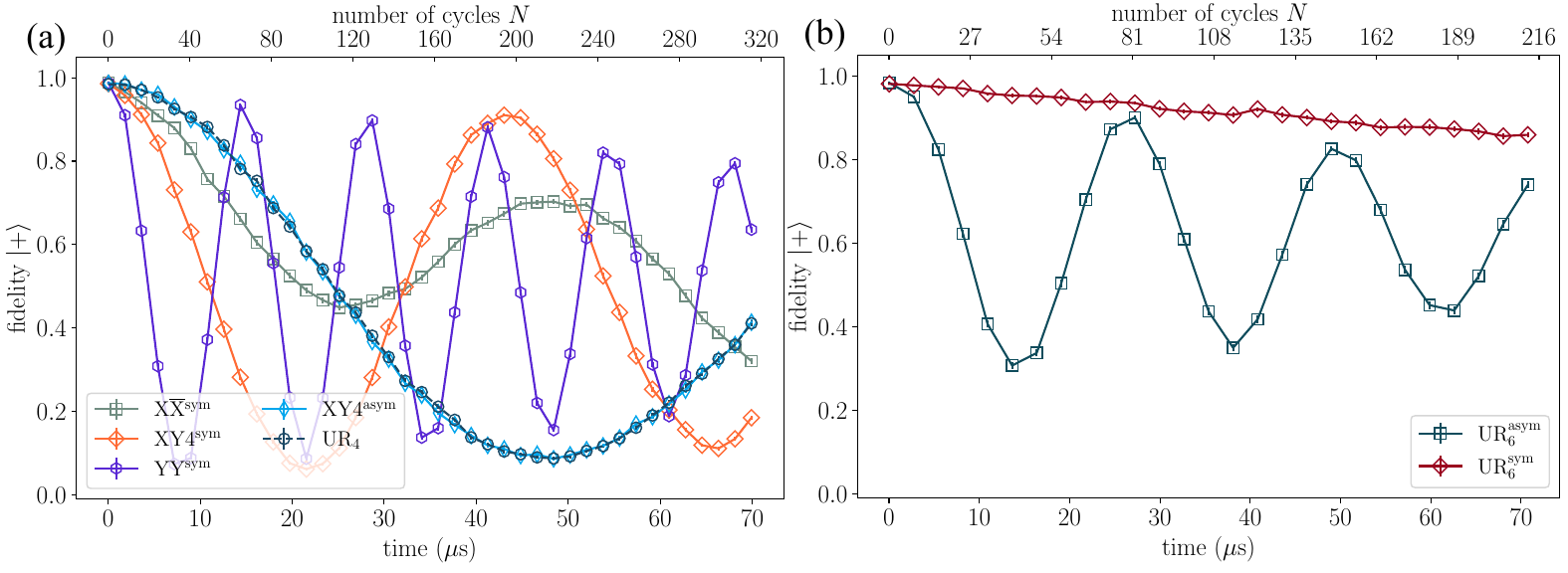}}
\caption{(a) The $\ket{+}$ state fidelity as a function of total sequence time (or number of sequence cycles; top axis), subject to the XY4$^{\text{sym}}$, XY4$^{\text{asym}}$, UR$_4$, $YY^{\text{sym}}$, and $X\overline{X}$ sequences applied to a single qubit (qubit 37) on the $\texttt{ibm\_sherbrooke}$ device. Data points for the two-pulse-long $YY^{\text{sym}}$ and $X\overline{X}$ sequences are shown for every second cycle (i.e., their total number of cycles is $640$). The XY4$^{\text{asym}}$ and UR$_4$ sequences exhibit nearly identical fidelity decay behavior, clearly distinct from that of the XY4$^{\text{sym}}$ sequence, confirming that the asymmetric $Y$ gates transform XY4 into the UR$_4$ sequence. All sequences shown exhibit oscillations. (b) The $\ket{+}$ state fidelity as a function of total sequence time (or number of sequence cycles; top axis), subject to UR$_6^{\text{sym}}$ and UR$_6^{\text{asym}}$ applied to qubit 89 on the $\texttt{ibm\_sherbrooke}$ device. Cycles corresponds to six-pulse sequences. UR$_6^{\text{sym}}$ drastically outperforms UR$_6^{\text{asym}}$.} 
\label{fig:three}
\end{figure*}

\begin{figure}[t]
\hspace{0cm}{\includegraphics[scale=.63]{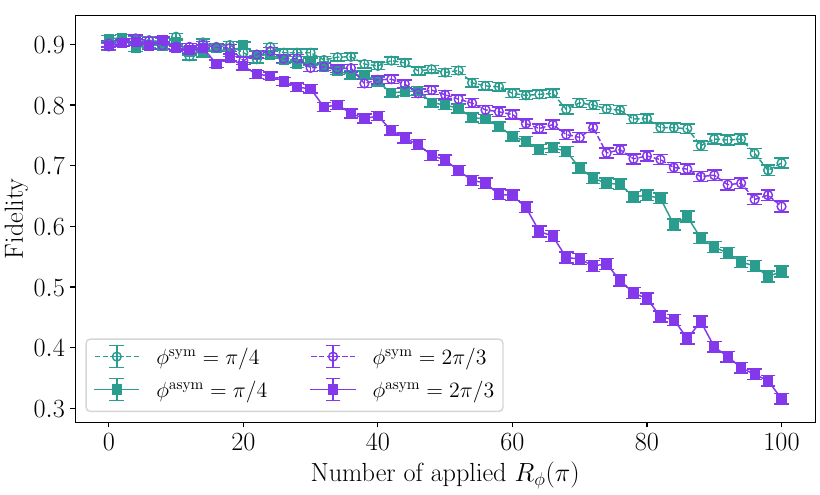}}
\caption{The impact of symmetric \textit{vs} asymmetric compilation on rotations in the $(x,y)$ plane. A GHZ$_3$ state is prepared and even numbers of symmetric or asymmetric rotations making an angle of $\phi=\pi/4$ or $\phi=2\pi/3$ with the $x$-axis are applied. The state is then unprepared and the fidelity is measured. The results show that the symmetric compilation consistently achieves higher fidelities compared to the asymmetric approach.}
\label{fig:algorithmic-GHZ3}
\end{figure}

\begin{figure*}[t]
\hspace{0cm}{\includegraphics[scale=.63]{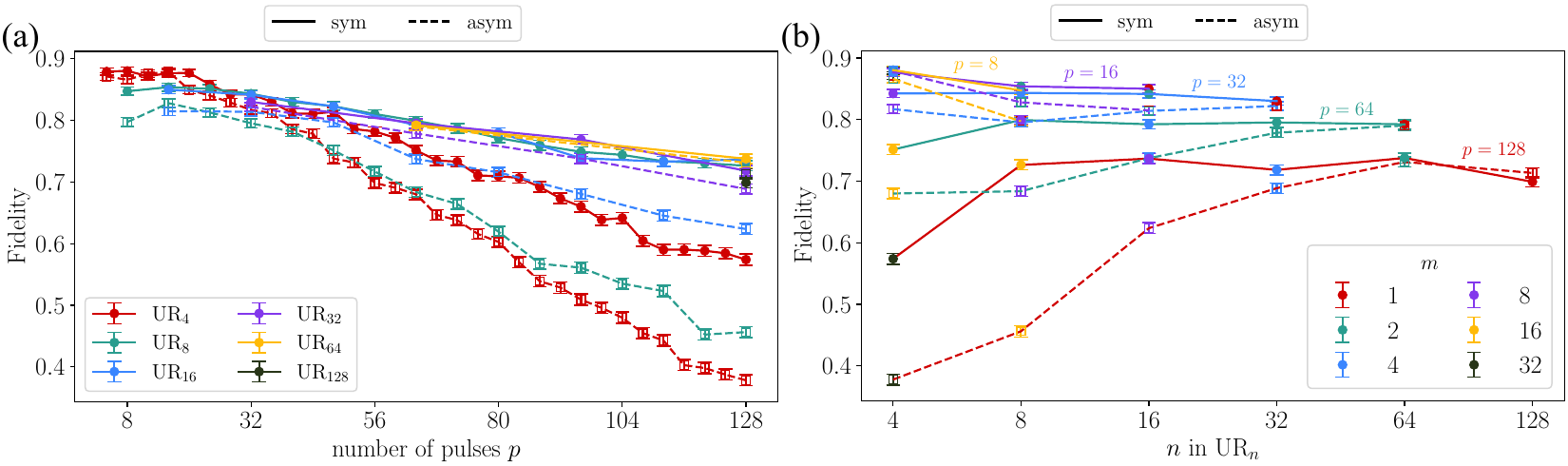}}
\caption{The fidelity of a three-qubit GHZ state subject to UR$_n$ sequences applied simultaneously to each qubit, using both symmetric and asymmetric compilations. In (a) we show the results as function of the total number of pulses $p$ (the total time elapsed is $p\tau$, where $\tau=56.8\,$ns). All UR$_n$ sequences are periodically repeated (e.g., UR$_4$ is repeated up to $32$ times), with the exception of UR$_{128}$, which is implemented only once. For $n\le 32$ we observe that symmetric compilation outperforms asymmetric compilation at all times. For $n=64$ and $128$, the two compilations are statistically indistinguishable. 
In (b) we plot (a subset of) the same data as a function of $n$ for different numbers of sequence repetitions $m$. Fixing $n$ and reading the results from top to bottom is equivalent to reading panel (a) from left to right. Reading the results from left to right along the connecting lines drawn gives the performance of UR$_n$ as a function of $n$ at fixed $p$ (total number of pulses). Either way, symmetric compilation outperforms asymmetric compilation. The results at constant $p$ also show that the performance of UR$_n$ deteriorates as a function of $n$ for the shorter sequences ($p\in \{8,16,32\}$), and improves with $n$ for the longer sequences ($p\in \{64,128\}$).}
\label{fig:other-URs}
\end{figure*}

\section{Experimental verification}
\label{sec:expt}

In this section, we report on various experiments to test our predictions about the role of \VZ gates in open quantum systems. We present both DD experiments and quantum circuits corresponding to GHZ state preparation.
The results illustrate the importance of the choice of symmetric compilation in the implementation of 
quantum circuits.

\subsection{DD sequence implementation using symmetric \textit{vs} asymmetric compilation}
\label{sec:expt-DD}

To test our prediction that when using $Y^{\text{asym}}$, XY4 is effectively the same as UR$_4$, \cref{fig:three}(a) presents the results of measuring the fidelity of the $\ket{+}$ state as a function of time for a variety of different pulse sequences, each of which is applied repeatedly. 
Specifically, we apply the following sequences to a single qubit on \texttt{ibm\_sherbrooke}: UR$_4$ using the symmetric definition for $\overline{X}$ given in \cref{eq:xbar}, and two versions of XY4 using the symmetric and asymmetric $Y$ gates. As shown in \cref{fig:three}(a), the UR$_4$ and XY4$^{\text{asym}}$ sequences are almost indistinguishable, as expected. In contrast, XY4$^{\text{sym}}$ is distinct.

It is important to note that Qiskit \cite{javadiabhari2024arxiv} natively compiles the $Y$ gate in the asymmetric form of \cref{eq:Yasym}. Therefore, caution is necessary when interpreting previously reported DD results involving transmon qubits that did not use the $Y^{\text{sym}}$ gate, including numerous studies involving the XY4 sequence. 

\cref{fig:three}(b) shows the performance of the UR$_6$ sequence: $X f_\tau R_{2\pi/3}(\pi) f_\tau X f_\tau X f_\tau R_{2\pi/3}(\pi)f_\tau X f_\tau$. For $R_{2\pi/3}(\pi)$, we used the asymmetric and symmetric decompositions given in \cref{eq:rphi-asym,eq:rphi-sym} in the sequences $\rm{UR}_6^{\rm{asym}}$ and $\rm{UR}_6^{\rm{sym}}$, respectively. Evidently, the compilation choice leads to two different sequences, with the symmetric option significantly outperforming the asymmetric one. This is a clear indication that the use of symmetric compilation is important not only for $Y$ or $\overline{X}$ gates, but also for rotations about arbitrary axes in the $(x,y)$ plane.

\subsection{Impact of symmetrically \textit{vs} asymmetrically compiled gates on GHZ state fidelity} 
\label{sec:GHZ-expt}

To test the impact of the choice of compilation beyond the single-qubit case, we now consider the problem of fidelity preservation of Greenberger–Horne–Zeilinger (GHZ) states. 

We prepare the three-qubit GHZ state (GHZ$_3$) and then simultaneously apply even numbers of $R_\phi(\pi)$ rotations with $\phi\in\{\pi/4,\pi/3\}$ to all qubits, implemented according to \cref{eq:rphi-asym,eq:rphi-sym}. Note that GHZ state-preparation itself does not involve gates affected by symmetric \textit{vs} asymmetric compilation. We then unprepare the GHZ$_3$ state and define the fidelity as retrieving the $\ket{000}$ state. In \cref{fig:algorithmic-GHZ3} we show that this leads to a clear fidelity difference, with symmetric compilation outperforming asymmetric compilation. 

As another test, we prepare the GHZ$_3$ state, apply a certain number of repetitions of UR$_n$ simultaneously to all three qubits for different values of $n$, unprepare the GHZ$_3$ state, and measure each qubit in the computational basis. We note that pulse staggering is the better choice for crosstalk suppression~\cite{Zhou2023PRL,evert2024syncopateddynamicaldecouplingsuppressing,brown2024efficient}, but for our purposes the simultaneous sequence suffices. 

Each UR$_n$ sequence consists of $n$ pulses, and we consider $n\in\{4,8,16,32,64,128\}$. When we repeat a given UR$_n$ sequence $m$ times (``periodic DD''~\cite{Khodjasteh:2007zr}), the total number of pulses is $p=mn$. We use both the symmetric and asymmetric compilations for each UR$_n$ sequence.

The results in \cref{fig:other-URs} illustrate that symmetric compilation always yields improved performance. This effect is more significant for smaller $n$ and for larger total pulse number $p$.

\cref{fig:other-URs}(b) also presents the first experimental evidence for the theoretically expected improvement with $n$ at constant total pulse number~\cite{Genov2017PRL}. However, we find that the improvement is observable only for sufficiently long pulse sequences, i.e., $p\ge 64$. Interestingly, the improvement is significantly more pronounced for the asymmetric compilation. A potential explanation is that for low $n$, the asymmetric sequence deviates more from ideal UR$_n$ than the symmetric sequence, and the difference between the two becomes less pronounced with increasing $n$. For the symmetric compilation, we find that increasing $n$ has only a small effect, except from UR$_4$ to UR$_8$ when $p\ge 64$.

\subsection{Pulse interference}

\cref{fig:three}(a) also displays the $YY$ and $X\overline{X}$ sequences, constructed using the symmetric definitions given in \cref{eq:Ysym,eq:xbar}, respectively.
An unexpected feature observed in \cref{fig:three}(a) is that all five sequences shown (including the robust ones), exhibit oscillations, which typically arise from coherent errors. We hypothesize that this phenomenon is due to an interference effect between consecutive pulses, e.g., due to an impedance mismatch in the microwave control lines~\cite{gross2024characterizingcoherenterrorsusing}. This causes large enough errors to overwhelm the first-order robustness of UR$_4$.

\begin{figure}[t]
\hspace{0cm}{\includegraphics[scale=.43]{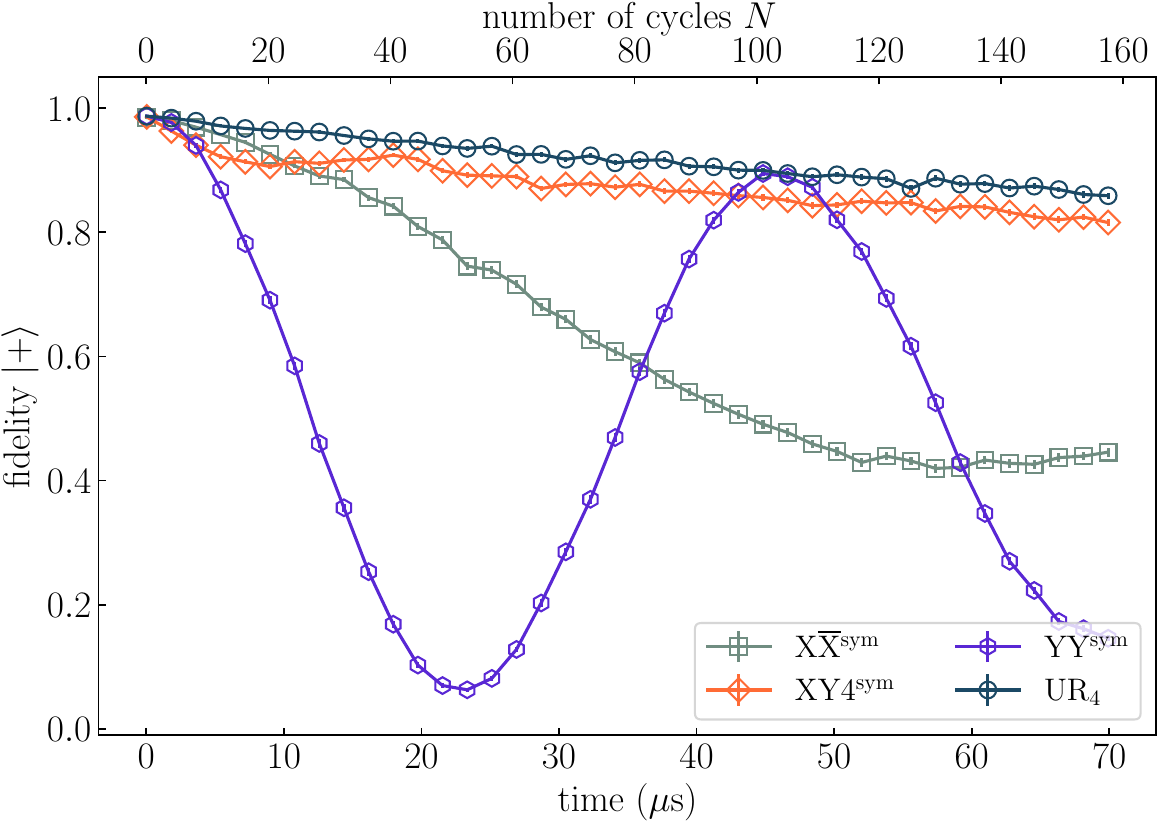}}
\caption{As in \cref{fig:three}(a) (except for the absence of XY4$^{\text{asym}}$) with the pulse interval doubled from $\tau=56.8$ ns to $2\tau=113.6$ ns. The oscillation periods of $YY^{\text{sym}}$ and $X\overline{X}$ increase significantly, and the difference between the now decaying XY4$^{\text{sym}}$ and UR$_4$ fidelities is nearly eliminated. } 
\label{fig:four}
\end{figure}

To test this hypothesis, we repeated the same experiments as shown in \cref{fig:three}(a) (except that we did not repeat XY4$^{\text{asym}}$ since we already established its equivalence with UR$_4$), but with an intentional delay added between consecutive pulses, thus doubling the pulse interval $\tau=56.8$ ns, defined as the time delay between the peaks of two consecutive pulses. The result, shown in \cref{fig:four}, is that the XY4$^{\text{sym}}$ and UR$_4$ sequences no longer oscillate, but exhibit simple decay. Moreover, the stark difference between the latter two sequences seen in \cref{fig:three}(a) has now almost disappeared. This is consistent with the observation that $Z$ (dephasing) errors and rapid (i.e., fast compared to the gate time) relaxation errors are the dominant error source in transmon qubits, so that sequences suppressing $X$ or $Y$ errors have little added benefit over sequences suppressing only $Z$ errors. 

The fact that the fidelities seen in \cref{fig:three}(a) are higher for intervals of $2\tau$ rather than $\tau$ also helps to explain why previous studies involving transmon qubits~\cite{Pokharel2018PRL,EzzellPRApp2021} have observed that, in contrast to the DD theory for ideal, zero width pulses (e.g., Ref.~\cite{UL:10}), the optimal pulse interval is not always the shortest possible (the same phenomenon has also been observed in other platforms, e.g., nuclear magnetic resonance~\cite{Peng:2011ly} and trapped ions~\cite{Morong:2023aa}). The pulse interference effect, with pulses applied consecutively with the minimum shortest possible pulse interval $\tau$, can introduce additional coherent errors that result in inferior DD performance even with sequences (such as UR$_4$) that are robust against small coherent errors. Our results confirm the conclusion of Ref.~\cite{EzzellPRApp2021} that it is essential to optimize the pulse interval for a given quantum processor, with the added insight that this optimization can reduce or (depending on the pulse sequence) even eliminate coherent errors due to pulse interference.

\begin{figure*}[t]
\hspace{0cm}{\includegraphics[width=0.98\linewidth]{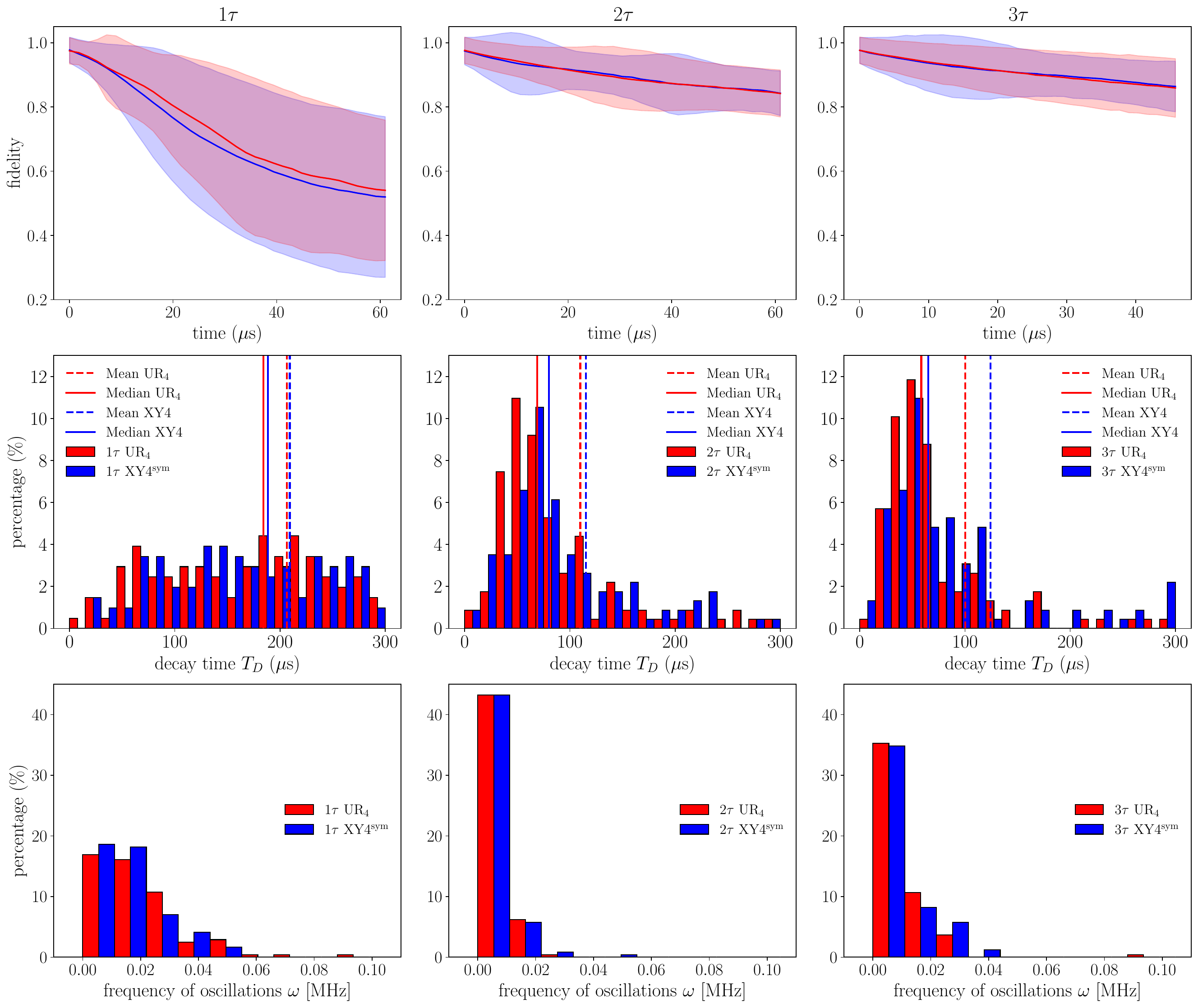}}
\caption{Fidelities of the XY4$^{\text{sym}}$ (blue) and UR$_4$ (red) sequences with different pulse intervals: $1\tau$, $2\tau$, and $3\tau$ (where $\tau = 56.8$ ns) for \texttt{ibm\_sherbrooke}. Top row: the corresponding fidelity means and standard deviations for $123$ out of the $127$ qubits for the indicated pulse intervals. For $1\tau$, the UR$_4$ sequence has a somewhat smaller decay rate. The decay rates decrease significantly at $2\tau$ and $3\tau$, as do their standard deviations. Middle row: the distribution of fitted decay times $T_D$ from fitting each of the $123$ decay curves to $a + b e^{-t/T_D}\cos(\omega t)$. The distribution is relatively broad at $1\tau$, and narrows significantly at $2\tau$ and $3\tau$. Both the mean and median XY4 decay times are slightly larger than UR$_4$'s for all pulse intervals, opposite from the top row result at $1\tau$. However, note that top row displays the mean at each time, which is different from the mean of the fits shown in the histograms. 
Bottom row: the  distribution of frequency of oscillations. Oscillations are relatively significant at $1\tau$, but as we increase the pulse interval to $2\tau$ and $3\tau$ the distributions peak around zero, indicating that oscillations are strongly suppressed. The difference in oscillation frequencies between the XY4 and UR$_4$ sequences is negligible.} 
\label{fig:five}
\end{figure*}

The reason we include the $X\overline{X}$ and $YY$ sequences in \cref{fig:three}(a) is that $X\overline{X}$ is susceptible to phase errors, while $YY$ is susceptible to rotation errors [\cref{eq:dth-dphi}], as discussed in detail in Ref.~\cite{tripathi2024-DB}. More specifically, \cref{fig:three,fig:four} shows that the two sequences exhibit oscillations for both the $\tau$ and $2\tau$ cases, with a period significantly shorter than that of the other sequences shown. This is consistent with the existence of single-pulse phase and rotation errors in addition to pulse interference errors. Doubling the pulse interval significantly increases
the oscillation period, as seen in \cref{fig:four}, but does not eliminate the oscillations. Moreover, we have checked (not shown) that further increasing the pulse interval to $3\tau$ has little effect on the $X\overline{X}$ and $YY$ fidelities, showing that coherent phase and rotation errors cannot be eliminated by controlling the pulse interference effect alone.

\cref{fig:four} displays results from a single qubit; to test whether the result of increasing the pulse interval is a statistically significant feature, we applied the XY4$^{\text{sym}}$ and UR$_4$ sequences to all $127$ qubits of the $\texttt{ibm\_sherbrooke}$ device, for pulse intervals of $\tau = 56.8$ns, $2\tau$, and $3\tau$. The results are shown in \cref{fig:five}, after removing four of the qubits whose measurements were inadvertently performed during a calibration cycle (qubits 20, 21, 56, 63). 

As seen in the top row, the oscillations exhibited by both XY4$^{\text{sym}}$ and UR$_4$ in the $1\tau$ case are significantly reduced for pulse intervals of $2\tau$ and $3\tau$. The time-averaged decay rate improves significantly when the pulse interval increases from $\tau$ to $2\tau$, showing that an optimized pulse interval (as opposed to the minimum interval) is preferable, consistent with the findings of Ref.~\cite{EzzellPRApp2021}.

In the middle row of \cref{fig:five}, we plot the distribution of decay times $T_D$ obtained by fitting the fidelities of the XY4$^{\text{sym}}$ and UR$_4$ sequences to $a + b e^{-t/T_D}\cos(\omega t)$. The distribution is quite broad for both XY4 and UR$_4$ at $1\tau$, and narrows considerably at $2\tau$ and $3\tau$, while shifting to the left and peaking at a slightly lower $T_D$ value for UR$_4$.

Finally, in the bottom row of \cref{fig:five}, we plot the distribution of the frequency of oscillations $\omega$ obtained by fitting the fidelities of the XY4$^{\text{sym}}$ and UR$_4$ sequences to the same function $a + b e^{-t/T_D}\cos(\omega t)$. Evidently, as we increase the pulse interval from $\tau$ to $2\tau$, the distribution moves toward smaller values, but broadens again from $2\tau$ to $3\tau$, suggesting that a $2\tau$ pulse interval is preferred.

Overall, we may conclude that the pulse interference effect is significant when pulses are applied back-to-back but is strongly suppressed by increasing the pulse interval. Optimizing the pulse interval can significantly increase the decay time while narrowing its distribution, and at the same time drastically reduce fidelity oscillations.

We note that pulse interference effects have been previously observed (e.g., Ref.~\cite{GustavssonPRL2013}). The key finding we report here is the ability to distinguish pulse interference from coherent pulse errors (such as phase and rotation errors) that robust sequences, such as XY4$^{\text{sym}}$ and UR$_4$, are designed to withstand. This distinction becomes apparent only after using the correct \VZ decomposition, as this allows us to implement the intended set of UR$_n$ sequences that are immune to specific coherent errors, but not to pulse interference.

\begin{figure*}[t]
\hspace{0cm}{\includegraphics[scale=.64]{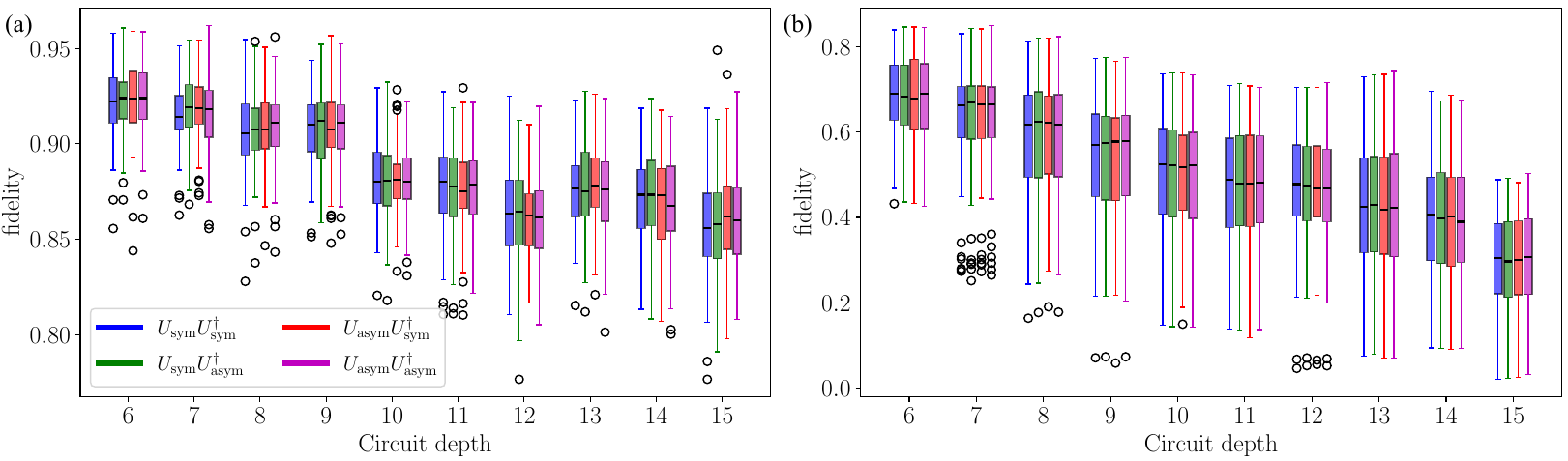}}
\caption{Box plot of the fidelity of $5$-qubit random circuits of various circuit depths. (a): \texttt{ibm\_marrakesh}, (b): \texttt{ibm\_sherbrooke}. Red lines are medians, boxes are the $25$'th to the $75$'th percentiles (quartiles), whiskers are $1.5$ times the interquartile range, and circles denote outliers. At each circuit depth, we generate $100$ random circuits distinguished only by the choice of symmetric or asymmetric compilation, as indicated by the legend. There is no statistically significant difference in fidelity between the four versions shown. 
The experiments reported in this plot used temporally balanced circuits (ASAP for $U$ and ALAP for $U^\dagger$) and bidirectional two-qubit gates.
}
\label{fig-rcs-1}
\end{figure*}

\section{Conclusion}
\label{sec:conc}

This work highlights the critical role of the \VZ gate and its interplay with the open system dynamics of quantum processors. We have demonstrated that a symmetric compilation of quantum gates with respect to the \VZ gate, e.g., the $Y$ and $\overline{X}$ gates, significantly improves the fidelity of these gates. In particular, it removes an undesired asymmetry between states in the $(x,y)$ plane of the Bloch sphere that is present when an asymmetric gate compilation is used instead. We have experimentally validated the advantage offered by a symmetric gate compilation using our in-house processor \texttt{MUNINN} as well as using the IBM cloud processor $\texttt{ibm\_sherbrooke}$, showing in particular the impact on commonly used DD sequences, as well as on GHZ state preservation.

Our findings highlight the need to carefully consider \VZ gate compilation in future studies, as well as the impact on previous studies that used asymmetric gate compilations. Specifically, we have shown that asymmetric compilation can lead to unexpected outcomes, such as fidelity asymmetries and incorrect implementations of DD sequences. This can result in misleading interpretations of earlier DD experiments. A case in point is that an asymmetric compilation of the $Y$ gate has the effect that the standard XY4 DD sequence is actually an implementation of the UR$_4$ sequence, which does not suppress any undesired $X$-type interactions. In contrast, symmetric compilations preserve the intended gate operations and result in a faithful implementation of the desired DD sequences. A related direction for future research is the impact of our results on twirling-based experiments, e.g., in the context of Probabilistic Error Cancellation~\cite{vandenBerg2023NaturePhys2023,KimNature2023}, where we expect VZ gate-based compilations based on real $Y$ gates to lead to more efficient suppression of off-diagonal terms in noise channels.

Going beyond DD and the $Y$ gate, we have also demonstrated that symmetric compilation of arbitrary $(x,y)$ plane rotations can affect algorithmic performance using the example of GHZ state preparation. In \cref{sec:algos} we further explore the effect on 
fidelity of random quantum circuits. 
Furthermore, we explored the impact of pulse interference, which can introduce coherent errors even in DD sequences that are designed to be robust to such errors. We have demonstrated that these effects can be mitigated by intentionally increasing the pulse interval, highlighting the importance of optimizing the pulse interval for a given quantum processor. These results explain earlier observations where robust sequences resulted in suboptimal performance; this effect can now be attributed to pulse interference effects.

Future studies may focus on refining gate compilation strategies and addressing pulse interference effects to further enhance the fidelity of quantum gates.

\begin{figure*}[t]
\hspace{0cm}{\includegraphics[scale=.64]{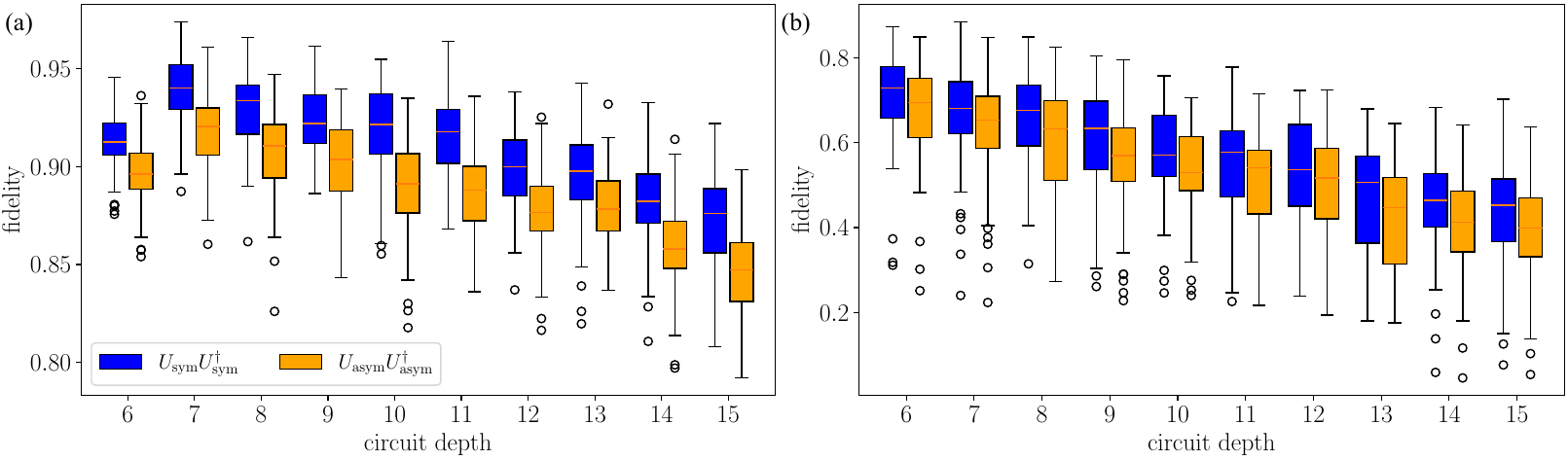}}
\caption{Box plot of the fidelity of $5$-qubit random circuits of various circuit depths, without temporal balance (ASAP for both $U$ and $U^\dagger$) and using only unidirectional two-qubit gates. (a): \texttt{ibm\_marrakesh}, (b): \texttt{ibm\_sherbrooke}. As in \cref{fig-rcs-1}, red lines are medians, boxes are the $25$'th to the $75$'th percentiles (quartiles), whiskers are $1.5$ times the interquartile range, and circles denote outliers.
At each circuit depth we have $100$ random circuits, each of which uses either all symmetric or all asymmetric compilation. The 
symmetric approach achieves consistently higher median and interquartile fidelities across all circuit depths.}
\label{fig-rcs-2}
\end{figure*}

\section*{Acknowledgement}
This material is based upon work supported by, or in part by, the National Science Foundation Quantum Leap Big Idea under Grant No. OMA-1936388, by the Defense Advanced Research Projects Agency under Agreement HR00112230006, by the Intelligence Advanced Research Projects Activity (IARPA) under the Entangled Logical Qubits program through Cooperative Agreement Number W911NF23-2-0216, and  by the U.S. Army Research Laboratory and the U.S. Army Research Office under contract/grant number W911NF2310255. The views, opinions and/or findings expressed are those of the author(s) and should not be interpreted as representing the official views or policies of the Department of Defense or the U.S. Government. In-house processor was fabricated and provided by the Superconducting Qubits at Lincoln Laboratory (SQUILL) Foundry at MIT Lincoln Laboratory, with funding from the Laboratory for Physical Sciences (LPS) Qubit Collaboratory. This research was conducted using IBM Quantum Systems provided through University of Southern California’s IBM Quantum Innovation Center. The views expressed are those of the authors and do not reflect the official policy or position of IBM or the IBM Quantum team.

\appendix

\section{Symmetric compilation in random quantum circuits}
\label{sec:algos}

Arbitrary rotations are often used in quantum algorithms, such as random circuit sampling~\cite{Zlokapa:2023aa}, variational quantum eigensolvers~\cite{Cerezo2021}, quantum simulation~\cite{Fauseweh:2024aa}, and the quantum Fourier transform~\cite{baumer2024quantumfouriertransformusing}. In this appendix we present various tests of the role of symmetric compilation in the setting of random quantum circuits. 

We prepare $5$-qubit random circuits of depths $6$ to $15$ (number of time steps). 
At each time step, we randomly assign single-qubit or two-qubit gates to some of the qubits, arranged in a linear chain. Our experiments use the \texttt{ibm\_sherbrooke} and \texttt{ibm\_marrakesh} QPUs; the former uses echo-cross resonance~\cite{Tripathi2019PRB} for its two-qubit gates, while the latter uses tunable couplers~\cite{Stehlik2021PRL}. We randomly select 2-qubit gates from the set $\{\text{CNOT, SWAP, CZ}\}$, applying them only to adjacent qubits. This ensures that long-range two-qubit gates are not broken into more two-qubit gates on other qubits. We can choose to apply random two-qubit gates in one of two ways: either in a unidirectional manner, where they are always indexed as \( q_i \rightarrow q_{i+1} \), or in a bidirectional manner, where they can be indexed as \( q_i \leftrightarrow q_{i+1} \). Here, \( i \) represents the index of the qubits in our five-qubit linear chain.

Single-qubit gates are chosen uniformly from the set $\{X, H, R_\phi\}$, where $H$ is the Hadamard gate and $R_\phi(\pi)$ is a rotation with a random angle $\phi$ such that $0 \leq \phi \leq 2\pi$. At least one of the five qubits always receives a gate at each time step, while the assignment of gates to the remaining four qubits is determined randomly. Thus, the total number of gates applied at each time step is dynamic and random. For each such random circuit $U$, we then run $U^\dag$ and measure each qubit in the computational basis; the fidelity is computed as the fraction of all-zero bitstrings. We generate $100$ different random circuits for each circuit depth, and run each random circuit once using the asymmetric choice of $R_\phi(\pi)$ [\cref{eq:rphi-asym}] and once using the symmetric choice [\cref{eq:rphi-sym}]. In Qiskit, the default scheduling options for a given circuit are either ASAP (As Soon As Possible) or ALAP (As Late As Possible)~\cite{javadiabhari2024arxiv}. Due to the significant difference in duration between single- and two-qubit gates, either option alone for both $U$ and $U^\dagger$ results in a temporal imbalance in the $UU^\dagger$ circuits. To ensure temporal symmetry across the entire circuit, we schedule $U$ using ASAP and $U^\dagger$ using ALAP, then stitch the two circuits together. Finally, in order to demonstrate that the choice of compilation leads to different unitaries, we also perform the experiments $U_{\rm{sym}}U_{\rm{asym}}^\dagger$ and $U_{\rm{asym}}U_{\rm{sym}}^\dagger$. If the choice of compilation does not matter, then these experiments should be identical to $U_{\rm{sym}}U_{\rm{sym}}^\dagger$ and $U_{\rm{asym}}U_{\rm{asym}}^\dagger$. 

\cref{fig-rcs-1} presents our fidelity results for all four circuit types across both QPUs. For each circuit depth, the exact same circuit was executed in the four different versions indicated: $U_{\rm{sym}}U_{\rm{sym}}^\dagger$, $U_{\rm{sym}}U_{\rm{asym}}^\dagger$, $U_{\rm{asym}}U_{\rm{sym}}^\dagger$, and $U_{\rm{asym}}U_{\rm{asym}}^\dagger$. The variations in fidelity across the four different circuit types are not statistically significant. 
This is the case when we use bidirectional gates as well as enforce temporal balance using ASAP and ALAP.

In \cref{fig-rcs-2} we show the results of performing the same set of experiments, but with unidirectional gates and without respecting temporal balance, i.e., using ASAP for both $U$ and $U^\dagger$. In this case, we see that the $U_{\rm{sym}}U_{\rm{sym}}^\dagger$ circuits always yield a higher fidelity compared to $U_{\rm{asym}}U_{\rm{asym}}^\dagger$. 
We conjecture that this result might be due the fact that coherent errors build up and are not reversed in the the temporally imbalanced case (while they are, to some extent, in the balanced case), and the symmetric compilation is more robust against such errors. However, this cannot be the full explanation since, as shown in \cref{tab-fidelity-comparison}, restoring bidirectionality also restores the performance parity between symmetric and asymmetric compilation, even in the temporally imbalanced case.

To explain why also two-qubit gate directionality plays a role, we conjecture that, since compiling two-qubit gates in terms of the native gate set will include asymmetric single-qubit gates, unidirectional two-qubit gates create a directional asymmetry that once again favors enforced symmetric single-qubit gates. 
A full explanation is left for future work.

\begin{table}[h]
    \centering
    \renewcommand{\arraystretch}{1.8}
    \setlength{\arrayrulewidth}{1.2pt}
    \setlength{\tabcolsep}{12pt}
    \begin{tabular}{|c|c|c|}
        \hline
        \multicolumn{1}{|c|}{\textbf{Temporal}} & 
        \multicolumn{1}{c|}{\textbf{Two-qubit gates}} & 
        \multicolumn{1}{c|}{\textbf{Fidelity}} \\ 
        \hline
        \hline
        Imbalanced & Unidirectional & sym $>$ asym \\ 
        \hline
        Imbalanced & Bidirectional & sym $\approx$ asym \\ 
        \hline
        Balanced & Bidirectional & sym $\approx$ asym \\ 
        \hline
        Balanced & Unidirectional & sym $\approx$ asym \\ 
        \hline
    \end{tabular}
    \caption{Comparison of random circuit sampling fidelity results based on temporal balance and two-qubit gate directionality. Fidelity results supporting the first and third rows are shown in \cref{fig-rcs-2} and \cref{fig-rcs-1}, respectively. Results for the other two rows closely resemble \cref{fig-rcs-1} (not shown). }
    \label{tab-fidelity-comparison}
\end{table}

\bibliography{biblo}

\end{document}